\documentclass[12pt]{spieman}  
\usepackage{amsmath,amsfonts,amssymb}
\usepackage{graphicx}
\usepackage{setspace}
\usepackage{tocloft}

\title{Time Assignment System and Its Performance aboard the Hitomi Satellite}

\author[a,*]{Yukikatsu Terada}
\author[a]{Sunao Yamaguchi}
\author[a]{Shigenobu Sugimoto}
\author[a]{Taku Inoue}
\author[a]{Souhei Nakaya}
\author[a]{Maika Murakami}
\author[a]{Seiya Yabe}
\author[a]{Kenya Oshimizu}
\author[b]{Mina Ogawa}
\author[b]{Tadayasu Dotani}
\author[c]{Yoshitaka Ishisaki}
\author[d]{Kazuyo Mizushima}
\author[d]{Takashi Kominato}
\author[d]{Hiroaki Mine}
\author[d,e]{Hiroki Hihara}
\author[a]{Kaori Iwase}
\author[a,e]{Tomomi Kouzu}
\author[a,b]{Makoto S.\ Tashiro}
\author[b]{Chikara Natsukari}
\author[b]{Masanobu Ozaki}
\author[b]{Motohide Kokubun}
\author[b]{Tadayuki Takahashi}
\author[e]{Satoko Kawakami}
\author[e]{Masaru Kasahara}
\author[e]{Susumu Kumagai}
\author[f]{Lorella Angelini}
\author[f,g]{Michael Witthoeft}

\affil[a]{Graduate School of Science and Engineering, 
Saitama University, 255 Shimo-Ohkubo, Sakura, Saitama 338-8570, Japan}
\affil[b]{Institute of Space and Astronautical Science,
Japan Aerospace eXploration Agency,
3-1-1 Yoshinodai, Sagamihara, Kanagawa 229-8510, Japan}
\affil[c]{Department of Physics, Tokyo Metropolitan University, 1-1 Minami-Osawa, Hachioji, Tokyo 192-0397, Japan}
\affil[d]{NEC Corporation, 10, Nisshin-cho 1-chome, Fuchu, Tokyo 183-8501, Japan}
\affil[e]{NEC Space Technologies, Ltd., 10, Nisshin-cho 1-chome, Fuchu, Tokyo 183-8551, Japan}
\affil[f]{Exploration of the Universe Division, Code 660, NASA/GSFC, Greenbelt, MD 20771, USA}
\affil[g]{ADNET Systems, 6720 Rockledge Drive Suite 504, Bethesda, MD 20817, USA}

\cftpagenumbersoff{figure}
\cftpagenumbersoff{table} 
\begin{document} 
\maketitle

\begin{abstract}
Fast timing capability in X-ray observation of astrophysical objects is one of the key properties for the ASTRO-H (Hitomi) mission.
Absolute timing accuracies of 350 $\mu$s or 35 $\mu$s are required to achieve nominal scientific goals or to study fast variabilities of specific sources.
The satellite carries a GPS receiver to obtain accurate time information, which is distributed from the central onboard computer through the large and complex SpaceWire network.
The details on the time system on the hardware and software design are described.
In the distribution of the time information, the propagation delays and jitters affect the timing accuracy. Six other items identified within the timing system will also contribute to absolute time error. 
These error items have been measured and checked on ground to ensure the time error budgets meet the mission requirements.
The overall timing performance in combination with hardware performance, software algorithm, and the orbital determination accuracies, etc, under nominal conditions satisfies the mission requirements of $35 \mu$s. 
 This work demonstrates key points for space-use instruments in hardware and software designs and calibration measurements for fine timing accuracy on the order of microseconds for mid-sized satellites using the SpaceWire (IEEE1355) network.
\end{abstract}


{\noindent \footnotesize\textbf{*}Yukikatsu Terada,  \linkable{terada@phy.saitama-u.ac.jp} }

\begin{spacing}{2}   

\section{Timing Capability for the Hitomi Satellite}
\label{section:intro}
The X-ray astronomy mission Hitomi
was successfully launched in February 2016 
as the sixth in a series of Japanese X-ray 
observatory satellites~\cite{ASTROH}. 
It was an international mission led by JAXA in collaboration 
with USA, Canada, and European countries, 
aiming to observe astrophysical objects in the X-ray band from 0.5 to 600~keV.
The satellite carried three X-ray telescopes, 
Soft X-ray Spectrometer (SXS), Soft X-ray Imager (SXI), and Hard X-ray Imager (HXI)~\cite{SXS,SXI,HXI}, and one soft gamma-ray detector (SGD)~\cite{SGD}. 
One important scientific goal was 
to understand physical processes under the extreme environments of 
active and variable astrophysical objects 
such as black holes, neutron stars, binary stars, and active galactic nuclei. 
The most important requirement for the mission was its spectroscopic capabilities, that is, the resolving power of photon energies in the soft energy band below 10~keV achieved by the micro-calorimeter SXS~\cite{SXS} and wide-band coverage by the HXI~\cite{HXI} and SGD~\cite{SGD}.
However, fast timing capability is also a key performance requirement for understanding variable astrophysical objects in the X-ray band. 
According to typical time scales for variation in various X-ray sources~\cite{AHTIME}, an absolute timing accuracy of 350~$\mu$s covers most (but not all) of the scientific requirements for the Hitomi mission, and is achievable with conventional methods, as demonstrated by the previous X-ray mission Suzaku~\cite{HXD}.
Therefore, scientific requirements for the Hitomi timing system are defined as an absolute accuracy of 350~$\mu$s, a value that should be achieved even following a single-point on-orbit failure.
However, much higher timing accuracies may be necessary for some phenomena, such as X-ray emissions from millisecond pulsars, fast oscillations in low-mass X-ray binaries, and so on. This is one reason why Hitomi carried an onboard GPS receiver (GPSR; described in detail in Section~\ref{section:design_hardware}). Hitomi thus had the capability to achieve more accurate timing performance in nominal operations, and we defined best-effort goals as an absolute accuracy of $35~\mu$s, which is an order of magnitude over requirements but may not be achieved following a single-point failure.

Timing requirements are not applied to all mission instruments.
SXI is an accumulation-type detector CCD with typical exposures of 4.0~s in the nominal mode, so these requirements and goals ($350~\mu$s and $35~\mu$s) are not valid.
Similarly, the active shield detectors of SGD (hereinafter, SGD-SHIELD) are mainly used for anti-coincidence measurements for the main camera of SGD and are also used to monitor the full sky in the hard X-ray band to measure light curves of transient astrophysical objects at resolutions of a few tens of microseconds.
Note that requirements for the absolute timing accuracy of SGD-SHIELD is the same $350~\mu$s to perform cross-correlation with other full-sky monitoring instruments in space, but we do not apply the goal of $35~\mu$s to SGD-SHIELD.

The remainder of this paper is organized as follows. We summarize the system design for time assignments to mission data in hardware and software in Sections~\ref{section:design_hardware}
and~\ref{section:design_software}, respectively.
In Section~\ref{section:time_uncertainty}, 
we list items that affect timing accuracy and their error budgets.
Finally, we describe pre-launch characterization of the Hitomi timing system
in Section~\ref{section:verification}, 
and summarize the timing performance in Section~\ref{section:inorbit:summary}.

\section{Hardware Design of the Timing System}
\label{section:design_hardware}
\subsection{Design concept}
\label{section:design_hardware:timing}

Basic concepts for timing systems in Japanese X-ray astronomy satellites were established in the ASCA (The Advanced Satellite for Cosmology and Astrophysics~\cite{ASCA}) and Suzaku satellites. All payload instrument timings are synchronized to a single onboard clock, the clock values appear in the telemetry, and mission data times are assigned based on telemetry values in offline analyses on the ground. 
A monotonically increasing counter that tallies the single onboard clock is called the ``Time Indicator'' (TI).
Since the data handling system onboard Hitomi is well designed to achieve reliable communications with ground stations~\cite{ASTROH}, we decided to apply the basic concept of the timing system to the onboard telemetry-and-command system. 
Hitomi has sufficient redundancy, so no timing-specific component is introduced except for the GPSR.
To achieve 350~$\mu$s absolute timing accuracy even when the GPSR fails, the system can apply the same time assignment methods as Suzaku as a fallback, in which TI is calibrated by an on-ground atomic clock during communication between the spacecraft and the ground station.

\subsection{Distribution of onboard timing information }
\label{section:design_hardware:timing_distribution}
\begin{figure}[ht]
\centering
\includegraphics[width=0.75 \textwidth]{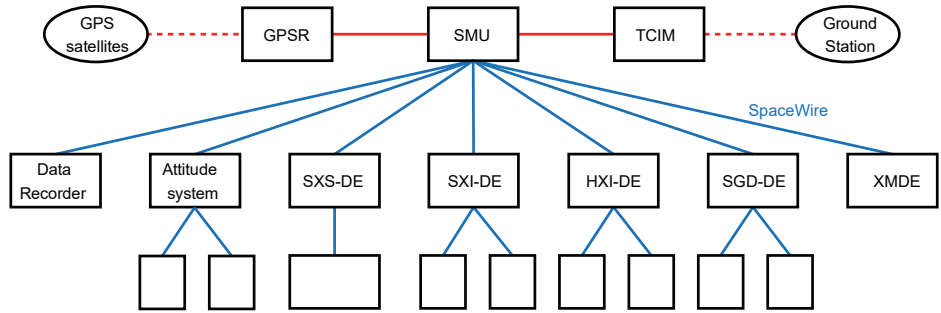}
\caption{A schematic diagram of the logical topology of the Hitomi network~\cite{AHnetwork}. Boxes represent components onboard the spacecraft and ellipses are GPS satellites or the ground station. Communication lines (in blue) are realized by SpaceWire.}
\label{fig:ahnetwork}
\end{figure}

\begin{figure}[ht]
\centering
\includegraphics[width=0.75 \textwidth]{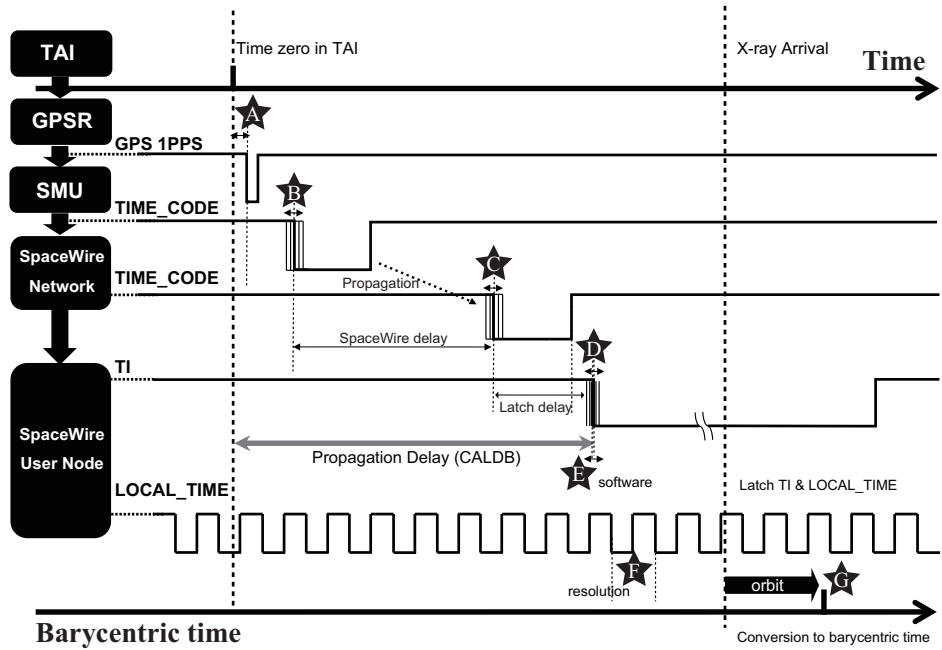}
\caption{Timing chart for distribution of timing information.
Error items in Table~\ref{tbl:astroh_time_error_budget} 
and calibration items are shown as stars and arrows, respectively.}
\label{fig:ahtiming}
\end{figure}

In the Hitomi timing system, the TI has a 38-bit length, covering time ranges from 15.625~ms (1/64~s) to about 136 years ($2^{38}/64$~s) in increments of 1/64~s, and is generated by a central onboard computer called the ``Satellite Management Unit (SMU)''~\cite{AHnetwork}, which is normally synchronized with International Atomic Time (TAI), provided by the GPSR.
Figure~\ref{fig:ahnetwork} shows a schematic of the logical topology of communication lines, and TI is distributed from SMU to payload instruments (SXS-DE, HXI-DE, etc. in the figure) through these communication lines.
Instruments receive TI from SMU every second, and the lower 32 bits of TI (hereinafter, L32TI) are latched and stored in a telemetry ``space packet'' to be sent to the data recorder.
The time in TAI when the space packet was generated can be reconstructed from the L32TI value in the space packet in offline analysis.
The left side of Fig.~\ref{fig:ahtiming} shows a schematic view of timing synchronization from TAI to the Hitomi system.

\subsection{Timing accuracy of synchronization between onboard instruments}
\label{section:design_hardware:spacewire}
Onboard telemetry and command communications
between instruments are implemented by ``SpaceWire,''
which is a standard network protocol based on IEEE~1355-1995~\cite{ SpaceWire:IEEE1355}.
The SpaceWire protocol itself defines only communication
methodologies in the physical and data-link layers, so 
the Hitomi mission adopts 
the remote memory access protocol (RMAP)~\cite{SpaceWire:RMAP}
as a session layer protocol.
To guarantee quality-of-service of the data transfer, 
the Hitomi system adopts the SpaceWire-D protocol~\cite{SpaceWire:D},
in which communication time is sliced into 1/64~s intervals 
by the highest-priority code in SpaceWire, 
which is called the ``TIME\_CODE''~\cite{SpaceWire:TIMECODE}.
For its logical topology, the Hitomi SpaceWire network has a tree structure 
(Fig.~\ref{fig:ahnetwork})
with its root node linked via one or more SpaceWire Routers to payload instruments
such as HXI-DE and SGD-DE,
which are called ``user nodes.''
The root node in the Hitomi network is the SMU 
(see Section~\ref{section:design_hardware:timing}),
which generates ``TIME\_CODE'' with well-controlled timing.
The 0th TIME\_CODE is always synchronized to time zero 
(values below the second are all zeros) of TAI.
In other words, TIME\_CODE is used for synchronization of 
all user nodes at an exact 64~Hz.

As described in Section~\ref{section:design_hardware:timing_distribution},
SMU distributes TI information to all user nodes.
The upper 32~bits of TI (covering above 1 s; hereinafter, U32TI) 
are distributed rather slowly (but within sub second)
from SMU to user nodes via the RMAP protocol, 
which is the same procedure as used in other telemetry-and-command transmission.
In contrast, the finer part of TI covering less than 1 s(6~bits) 
is quickly broadcasted within the SpaceWire network 
using the same high-priority code as is used for SpaceWire-D,
that is, the TIME\_CODE, which carries 6-bit information.
The Hitomi satellite carries a large number of SpaceWire nodes (up to 120),
and has main-and-redundant network paths, making the SpaceWire system
large and complex. Therefore, an understanding of the propagation 
delay and jitter of the SpaceWire TIME\_CODE is key 
to estimating the timing accuracy. 
As an example, the link rate of the Hitomi SpaceWire network at 50 or 20~MHz
results in delay and jitter on the order of about 1~$\mu$s
(for details, see Sections~\ref{section:verification:CD_timecode}),
which are non-negligible for timing goals of 35~$\mu$s 
(see Section~\ref{section:intro}).

\subsection{Assignment of photons at fine timing resolutions}
\label{section:design_hardware:localtime}
The arrival times of X-ray photons from astronomical objects
are determined at user nodes connecting to X-ray sensors
such as SXS, HXI, and SGD.
The time resolution of TI 
(15.625~ms; see Section~\ref{section:design_hardware:timing})
is insufficient for the timing goal of 35~$\mu$s 
(Section~\ref{section:intro}), 
although the timing accuracy of TI distribution is sufficient
on the order of about 1~$\mu$s (see Section~\ref{section:design_hardware:spacewire}).
Therefore, clocks with finer timing called ``LOCAL\_TIME'' 
are installed into each user node 
to refine resolutions to 5~$\mu$s, 61.0~$\mu$s, 25.6~$\mu$s, and 25.6~$\mu$s 
for SXS, SXI, HXI, and SGD, respectively, 
but with shorter time coverage than TI, 
as summarized in Table~\ref{tbl:summary_local_time}.
These time resolution values are within the requirement for $350~\mu$s accuracy for SXS, HXI, and SGD, and for a resolution of a few tens of milliseconds for SGD-SHIELD (see Section~\ref{section:intro}).
The time resolution for SXS is also sufficient for the best-effort goal of $35~\mu$s accuracy. 
The time resolutions of the HXI and the SGD ($25.6~\mu$s) may seem to be a comparably larger fraction of the timing goal ($35~\mu$s).
However, according to numerical studies of realistic scientific cases for determining the coherent periods of periodic X-ray signals from neutron-star pulsars, these instruments have sufficient performance to determine the period of $\sim 100$~ms at about less than 2~$\mu$s errors even from dimmer neutron stars below 1/1,000 of the X-ray flux of the Crab pulsar.

\begin{table}[htb]
\renewcommand{\arraystretch}{1.3}
\caption{Summary of LOCAL\_TIME counters}
\label{tbl:summary_local_time}
\centering
\begin{tabular}{lrr}
\hline 
Instrument & Bit length & Time resolution \\
\hline 
SXS        & 28-bits    &    5~$\mu$s \\
SXI        & 32-bits    & 61.0~$\mu$s \\
HXI        & 32-bits    & 25.6~$\mu$s \\
SGD        & 32-bits    & 25.6~$\mu$s \\
SGD-SHIELD & 32-bits    &   16 ms \\
\hline 
\end{tabular}
\end{table}

Since a phase-locked loop to the TI counter requires the hardware resources at each user node, the LOCAL\_TIME counters on Hitomi user node, except for the SXI, are implemented with free-run clocks to reduce the hardware resources.
In other words, LOCAL\_TIMEs are not synchronized to TI
and thus should be calibrated by TI with offline 
software (for details, see Section~\ref{section:design_software}).
Note that the LOCAL\_TIME for the SXI is synchronized to L32TI on every arrival of a TIME\_CODE at SXI-DE within $10~\mu$s accuracy, so such offline calculations are not required for SXI.
Therefore, SXS, HXI, and SGD user nodes
periodically (every 1 or 4~s) latch
LOCAL\_TIME and U32TI values simultaneously at TIME\_CODE = 0 
(i.e., U32TI time zero) to generate a lookup table 
between LOCAL\_TIME and TI.
This lookup table for each user node is periodically stored 
in a housekeeping space packet, with which 
the offline software recognizes the relation between
user node LOCAL\_TIMEs and TI.

When a user node acquires an X-ray event, 
the LOCAL\_TIME at detection is latched and stored in an telemetry message. 
Such information is gathered for multiple events into a space packet with a corresponding TI value. 
In summary, 
the time evolution of the relation between the LOCAL\_TIME 
and the TI is stored in housekeeping telemetry,
and the LOCAL\_TIME at X-ray detection and 
the TI value when a space packet containing the detection is generated
are stored in the event telemetry.
These three pieces of information in the telemetry are used in 
calculations of the photon-arrival time by the offline software.

\section{Software Design of Offline Time Assignment}
\label{section:design_software}
\subsection{Process flow}
\label{section:design_software:overview}

When all observation telemetry is ready on the ground, an automatic pipeline process starts converting from raw telemetry data into the Flexible Image Transport System (FITS) format, a standard format for astrophysical use~\cite{2001A&A...376..359H}. 
The process then calculates calibrated physical values such as time, pulse height invariant (photon energy), and coordinates from the raw telemetry values using the calibration database, and writes them to FITS output files. Since Hitomi is a "public observatory,"
the availability of such analysis software and calibration information for payload instruments is very important. Thus, all Hitomi software and calibration databases are published in the ``HEAsoft'' package and ``CALDB'' database, respectively, which are released to the public via NASA GSFC.

\begin{figure}[ht]
\centering
\vspace*{-0.1cm}
\includegraphics[width=0.75 \textwidth]{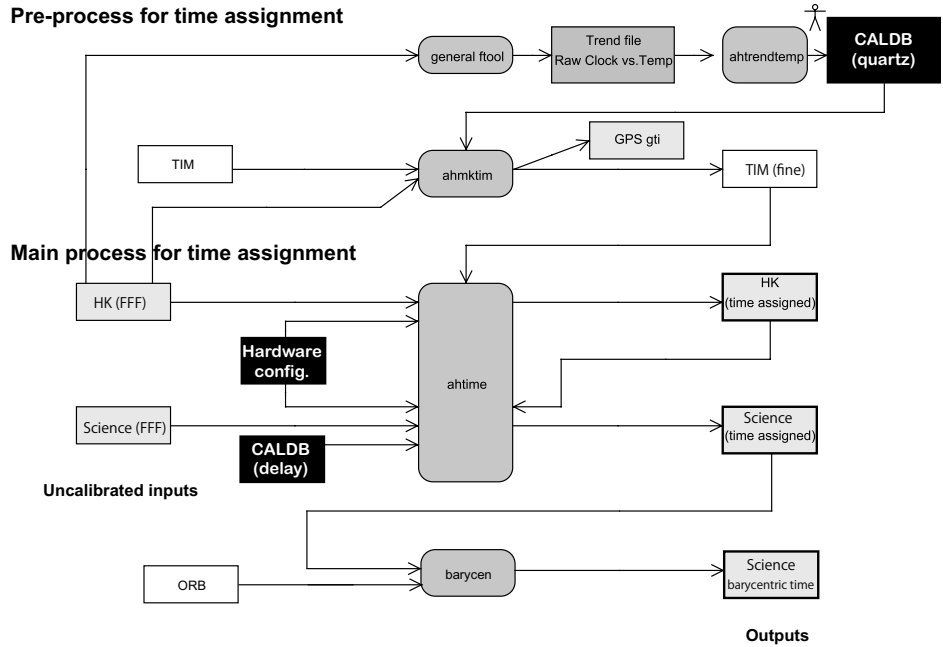}
\caption{UML chart for offline time assignment processes. 
Data and databases are shown in boxes; software is shown in rounded boxes.}
\label{fig:ahsoft}
\end{figure}

Time assignment tasks are applied to raw telemetry in a pipeline process after conversion to the FITS format.
Time calculations are performed one-by-one for all FITS housekeeping and science event files. 
The basic design concept for software time assignment tasks is common across all instruments. 
Information from individual payload instruments is stored in the CALDB files.
Figure~\ref{fig:ahsoft} shows an overall flow of the timing process as a universal modeling language (UML) chart.

The main process of the time assignment flow (Fig.~\ref{fig:ahsoft}) is provided by the main timing task {\it ahtime}, which calculates times for both housekeeping and science event files.
The algorithm for time assignment varies by file type (housekeeping or science event), by instrument, and by their operational mode, which can be recognized by FITS header keywords for timing tasks.
Individual information for {\it ahtime} is categorized as hardware configuration or propagation delay time from GPSR to SpaceWire nodes, and is stored in the two CALDB files shown as black boxes in Fig.~\ref{fig:ahsoft}. The algorithm is described in detail in Section~\ref{section:design_software:ahtime}.
The resulting time in TAI is converted into the terrestrial time (TT) values, whose origin is 2014-01-01 00:00:00~UTC, and stored into the TIME column in the FITS file by {\it ahtime}.

Pre-processing of the time assignment flow (Fig.~\ref{fig:ahsoft}), which is performed before the main process described above, is performed to handle the GPSR failure mode. 
GPS information can be commonly treated among all telemetry and affects only the relation between Hitomi TI and the corresponding TAI value. 
This relation is discontinuously measured during communications between the spacecraft and the ground station (as described in Section~\ref{section:design_hardware:timing}) and is stored in the TIM FITS file shown in Fig.~\ref{fig:ahsoft}.
Process by the {\it ahmktim} tool provides a continuous relation between the TI and TAI regarding the GPS receiving status. A table refined by {\it ahmktim} is stored in the TIM-fine file, which is used for individual time assignment by {\it ahtime} in the main process. 
The detailed algorithm of {\it ahmktim} is described later in Section~\ref{section:design_software:ahmktim_ahtrendtemp}, and utilizes properties of the quartz clock onboard SMU.
A temperature dependence of a clock frequency of this quartz is stored in CALDB (quartz), indicated by the black box in Fig.~\ref{fig:ahsoft}, and is scheduled for monthly onboard calibrations by another pre-process (the {\it ahtrendtemp} tool).

The barycentric dynamical time (TDB; the photon arrival time converted from the celestial sky position for light-travel times between the spacecraft and the center of the gravity of the solar system) is required for further astrophysical uses, for example, when matching the arrival time of photons in X-ray and radio observations~\cite{HITOMI_Crab_GRP}. A barycentric correction tool ({\it barycen}) is provided as an analysis tool, which is not applied in the pipeline process, because target positions for calculation must be set by users. 
The tool {\it barycen} is developed to be a mission-independent tool in the HEAsoft package, which rely on the data formats containing specific information in a standard way, expecting that the time pre-computed is stored in the TIME column with standard FITS keywords to identify the time system. Similarly it is expecting that the orbit is written in a file with well-defined quantities (position and velocity) written in standard column names and units.
Historically, the barycentric correction tools were individually developed for previous missions including ASCA and Suzaku~\cite{HXD}, mainly because the orbital information required by the tool is stored in a different format.
The engine of the calculation of the barycentric correction in {\it barycen} is a well trained algorithm derived as a standard routine for RXTE (Rossi X-ray Timing Explorer) satellite and after used by Chandra, Swift, Suzaku, NuSTAR (Nuclear Spectroscopic Telescope Array) and Fermi missions.
The calculation is performed in the two steps, following the same procedure used for the Suzaku satellite:
1) conversion from the TT value into the TDB value, considering the movement of the Earth under the gravitational potential of the Sun,
2) correction of the time delay of the light travel time between the spacecraft and the solar system barycenter, where the geometrical position of the spacecraft is identified by the orbital elements in the orbit file, considering the general relativity effects by the Sun and planets. 
The tool supports the solar system ephemerides of JPL-DE200.
After the conversion, the FITS header keyword TIMEREF is changed from 'LOCAL' to 'SOLARSYSTEM.'
The largest shifts (about 8 minutes maximum) occur in step 2, but this calculation is well established.
The largest systematic error arises in step 1, from positional accuracy for the orbital element of the spacecraft. Such systematic errors are discussed in Section~\ref{section:time_uncertainty}.

\subsection{ Algorithm for the time assignment process {\it ahtime}}
\label{section:design_software:ahtime}

Calculations at the main processing stage using {\it ahtime} are simple.
In calculating times for the housekeeping FITS files, the TAI system time is calculated from the input of L32TI, which is one value per one row in the FITS file using the TIM file (i.e., the relation between L32TI and TAI, as defined in Section~\ref{section:design_software:overview}).
The {\it ahtime} process takes from the TIM file two points before and two after the input L32TI and linearly interpolates them to calculate the corresponding TAI value. 
Since the TIME column in the Hitomi FITS file is defined as the number of seconds from 2014-01-01 00:00:00~UTC, whereas the origin for TAI is 1980-01-06 00:00:19~UTC, the calculated TAI is converted to the ASTRO-H time system by subtracting the offset of the two origins; i.e., the ASTRO-H time in TT is ${\rm TI}-1,072,569,616$ sec, where the origin of TI is ${\rm TAI}-19$ sec.

When calculating the TIME of science event FITS files from payload instruments using {\it ahtime}, LOCAL\_TIME counters, described in Section~\ref{section:design_hardware:localtime}, are also considered.
Since LOCAL\_TIME represents a finer timing when a photon is detected by the instrument, the TIME in the science event FITS file is defined at the event detection timing at finer time resolutions.
The calculation contains the two lookup processes described above.
The first lookup process uses the relation between LOCAL\_TIME and U32TI monitored in the housekeeping file of the instrument to calculate the L32TI corresponding to the input LOCAL\_TIME.
The second lookup process is the same procedure as that in the housekeeping data, from L32TI to TAI using the TIM file.
Since the propagation delay of TAI information from GPSR to the instrument's SpaceWire node is not negligible as compared with the time resolution (see Section~\ref{section:design_hardware:spacewire}), the final step in time assignment for science event files by {\it ahtime} is to add the propagation delay to the TAI value obtained above. As a result, 
the {\it ahtime} calculation retrieves the time in TAI with finer resolution than TI at the X-ray detection timing by each event, which can be converted to the ASTRO-H time system as was done for the housekeeping data.

To perform the above process, individual instrument information is stored in the CALDB files (black boxes in Fig.~\ref{fig:ahsoft}) and is used by {\it ahtime}. The stored information is
1) the CALDB hardware configuration, which describes hardware configuration parameters for LOCAL\_TIME as summarized in Table~\ref{tbl:summary_local_time} and the attribute names in the telemetries of instrument timing counters,
and
2) the CALDB propagation delay, which describes the propagation delay time of TAI information from GPSR to the SpaceWire user node via SMU and the SpaceWire network (Section~\ref{section:design_hardware:spacewire}).

\subsection{Treatment of GPS lock-off data, {\it ahmktim} and {\it ahtrendtemp}}
\label{section:design_software:ahmktim_ahtrendtemp}

As described in Section~\ref{section:design_software:overview}, pre-processing of the time assignment flow in Fig.~\ref{fig:ahsoft} checks the GPSR status to calculate the relation between TI and TAI, to support the fallback case in the event of GPSR failure as required in Section~\ref{section:intro}.

The GPSR onboard Hitomi is designed to always capture several GPS satellites on orbit (hereinafter, we call this GPS-ON mode), so cases where GPSR detects no GPS carry signal (hereinafter, GPS-OFF mode) should be rare.
In the design policy for the Hitomi spacecraft, the data acquisition system should be robust for any single-point failure, such as a completely GPS-OFF mode.
At the same time, the SpaceWire-D protocol used in the Hitomi data acquisition system (see Section~\ref{section:design_hardware:spacewire}) requires that the TIME\_CODE should not jump and that the time interval between TIME\_CODEs (time slot in SpaceWire-D; 16~ms) should be stable to within a few hundred microseconds.
The SMU is thus designed to follow the above requirements.
The TI is always synchronized to the TAI provided by the GPSR in GPS-ON mode and, even in GPS-OFF mode, the SMU keeps generating the TI using the free-run clock onboard SMU and keeps distributing TIME\_CODEs.
When the GPS-OFF mode switches to GPS-ON mode, the SMU starts synchronization of TI to TAI, which takes a few to few tens of seconds as a transition mode, and then the TI is synchronized to TAI again in GPS-ON mode.
The transition mode normally happens at the beginning of operations (such as initial on-orbit operations), but it never happens when GPS-OFF mode continues forever.

\begin{figure}[ht]
\centering
\includegraphics[width=0.75 \textwidth]{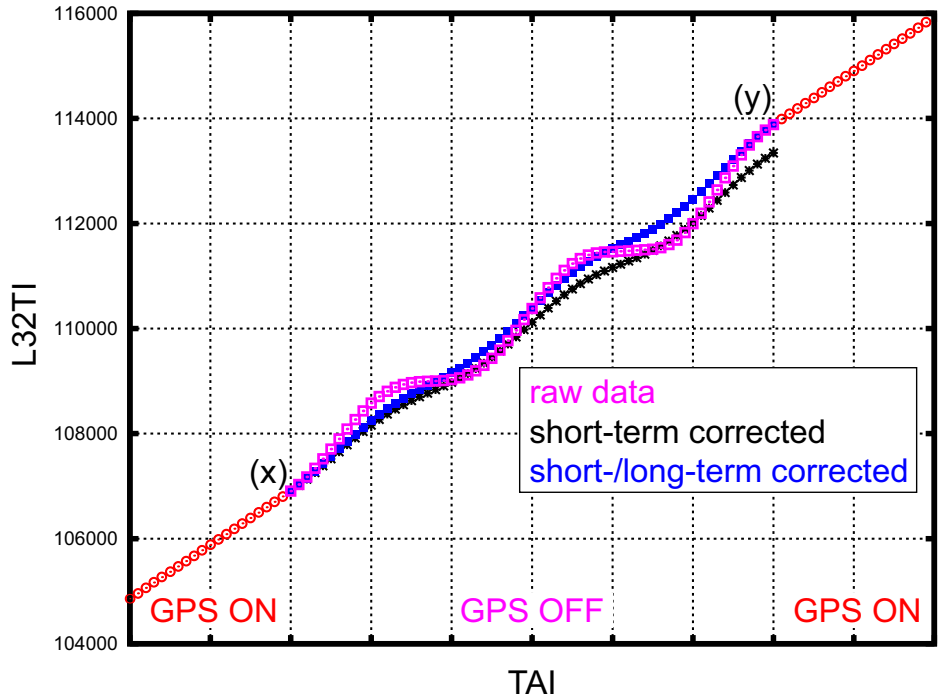}
\caption{Schematic view of calculating the relation between TAI and L32TI during GPS-OFF mode by {\it ahmktim}. Raw data obtained during GPS-ON and GPS-OFF modes are shown in red and magenta, respectively. Black points are intermediate data in the {\it ahmktim} calculations, where short-term temperature drifts were corrected. Blue data points are final task outputs, whose long-term shifts are corrected to match the final point (y).}
\label{fig:ahsoft_ahmktim}
\end{figure}

As briefly described in Section~\ref{section:design_software:overview}, the {\it ahmktim} process refines the TIM file, which describes the continuous relation between the TI and TAI, considering the GPSR status.
This relation is simple in normal GPS-ON mode, because the TI is synchronized to TAI. On the other hand, the calculation during GPS-OFF mode has two-step corrections of GPS-OFF data as schematically shown in Fig.~\ref{fig:ahsoft_ahmktim} when the GPSR status changes from GPS-ON to GPS-OFF, and then back to GPS-ON mode.
The raw data obtained during GPS-ON and GPS-OFF modes are shown in red and magenta, respectively, and a wavy trend can be seen during the GPS-OFF mode from TAI = (x) to (y).
Such fluctuations in TI are generated by a free-run clock on the SMU whose frequency shifts with temperature drift.
As the first step of GPS-OFF data correction, such short-term trends are corrected by the temperature of the SMU quartz, as monitored in the housekeeping data using the CALDB (quartz) file defined in Section~\ref{section:design_software:overview}, shown as black points in Fig.~\ref{fig:ahsoft_ahmktim}.
In the second correction step for GPS-OFF data, the red points during GPS-ON mode are used as anchor points where the TI is synchronized to TAI, and to adjust the beginning point (x) and the end point (y) of the GPS-OFF data to the two anchor points.
In the case of Fig.~\ref{fig:ahsoft_ahmktim}, the beginning point (x) is already connected to the last point of the former GPS-ON data, but a tentative jump between black and red points can be seen at (y), which is then corrected as a long-term correction. Finally, a more stable trend than the original data (magenta) is obtained as the blue points in Fig.~\ref{fig:ahsoft_ahmktim}.
The red and blue points are output from {\it ahmktim}.
Note that if no anchor points exist during the observation like in the case of complete GPS-OFF mode under permanent GPSR failure, {\it ahmktim} searches for other anchor points from the original TIM file, which describes the relation between TI and TAI as discontinuously measured during ground contacts.
This function provides a fallback plan that successfully worked for {\it Suzaku}.

Another pre-processing of time assignments in Fig.~\ref{fig:ahsoft} records the temperature dependence of the SMU quartz clock in the CALDB (quartz) file as measured on orbit. 
Information obtained on orbit is limited, so detailed measurements of the temperature dependence of the SMU quartz are performed on the ground, the results of which are described in Section~\ref{section:verification:suzaku}.
The SMU measures the frequency of the onboard quartz clock, referring to the GPS signals from GPSR. This function can be activated by an operation command during GPS-ON mode, but note that the measurement does not work during GPS-OFF mode.
Such operations are scheduled once per month on orbit.
Therefore, pre-processing by {\it ahtrendtemp} is scheduled to be triggered monthly.
The {\it ahtrendtemp} process picks up temperature and self-measured frequency information of the SMU quartz from housekeeping telemetry, and generates the relation between temperature and frequency. 
In this process, same temperature may appear many times, and thus the {\it ahtrendtemp} process calculates the average of the frequency values within the defined temperature bins. 
After the averaging operation, the results are stored in an extension of the CALDB file for the quartz frequency.

\section{Management of the Timing Uncertainties}
\label{section:time_uncertainty}
\subsection{Items on timing uncertainties}
\label{section:time_uncertainty:items}

Under the hardware and software designs for the timing system (see Sections~\ref{section:design_hardware} and \ref{section:design_software}), timing performance should meet the requirements for the timing system, namely, 350 or 35~$\mu$s in absolute time as a basic requirement or a mission goal, respectively (see Section~\ref{section:intro}).
To more precisely control overall uncertainty in the timing system, seven items are identified as those which may reduce timing accuracy. 
These items are indicated by stars in the timing chart of the Hitomi timing system, shown in Fig.~\ref{fig:ahtiming}. 
Table~\ref{tbl:astroh_time_error_budget} gives a summary of these items.

\begin{table*}[tbh]
\caption{Error budget in time assignments.}
\label{tbl:astroh_time_error_budget}
\centering
\begin{tabular}{lllc}
\hline 
ID & Component & Error Items &Error Budget\\
\hline 
A
& GPSR
& Jitter between TAI \& GPSR timing signal
& $<0.02~\mu$s (GPS-ON)\\
B 
& SMU
& Jitter between GPSR timing \& TIME\_CODE
& $<0.5~\mu$s (GPS-ON)\\
&
& 
& $<270~\mu$s (GPS-OFF)\\
C
& SpaceWire network
& Jitter between TIME\_CODE at SMU \& User node
& $<2.0~\mu$s\\
D
& SpaceWire user node
& Jitter between TIME\_CODE \& reconstructed TI
& $<1.0~\mu$s\\
E
& Software {\it ahtime}
& Uncertainty in reconstruction of TI
& $<1.0~\mu$s\\
F
& SpaceWire user node
& Resolution of LOCAL\_TIME 
& $\leq 25.6~\mu$s\\
G
& Ground system
& Accuracy of orbital elements
& $<3.0~\mu$s (= 1 km)\\
\hline 
\end{tabular}
\end{table*}

Under propagation of TAI timing information to SpaceWire user nodes (Fig.~\ref{fig:ahtiming}), timing delays are stored in the CALDB (delay) file and corrected by the offline timing tool {\it ahtime} (see Section~\ref{section:design_software:ahtime}), but jitter is treated as timing uncertainty items~{\bf A}, {\bf B}, {\bf C}, and {\bf D}.
In the first step, GPSR tries to synchronize the timing output signal (1~PPS; pulse per second) to the timing origin of TAI under GPS-ON mode (see Section~\ref{section:design_software:ahmktim_ahtrendtemp}). Item~{\bf A} is timing jitter between TAI and the timing output of GPSR.
Next, the SMU generates TI (Section~\ref{section:design_hardware}) using the timing signal from GPSR or the SMU quartz while in GPS-ON or GPS-OFF mode, respectively (see Section~\ref{section:design_software:ahmktim_ahtrendtemp}), and sends the TIME\_CODE as the lower 6 bits of the TI to the SpaceWire network. 
Item~{\bf B} is the timing jitter between the input and output of the SMU, that is, the timing signal from GPSR and the TIME\_CODE, respectively. 
Third, the TIME\_CODE is distributed via the SpaceWire network, and user nodes reconstruct the TI from TIME\_CODE and the upper 32 bits of the TI obtained via RMAP (see Section~\ref{section:design_hardware}). Item {\bf C} is the jitter of the timing of TIME\_CODE between, before, and after propagation via the SpaceWire network, and item~{\bf D} is jitter in synchronization of TIME\_CODE to the TI reconstructed on the user node.

Item~{\bf E} is systematic error in the interpolation of TIs by the timing task {\it ahtime} (see Section~\ref{section:design_software:ahtime}), and item~{\bf F} is the time resolution of instruments as summarized in Table~\ref{tbl:summary_local_time}.
Item~{\bf G}, which is accuracy of the orbital determination of the spacecraft, is provided for users who require barycentric time in their analyses. The barycentric time is calculated by the tool {\it barycen} (see Section~\ref{section:design_software:overview}), which requires the target position and the orbital elements of the spacecraft. 

\subsection{Error budget for timing uncertainties}
\label{section:time_uncertainty:err_budget}
To control and maintain overall timing uncertainties, the error budgets for all items identified in Section~\ref{section:time_uncertainty:items} are defined as in Table~\ref{tbl:astroh_time_error_budget}.
As defined in Section~\ref{section:intro}, the requirements and goals for SXS, HXI, and SGD are $350~\mu$s and $35~\mu$s, and item~{\bf F} is defined in Table~\ref{tbl:summary_local_time} ($5~\mu$s for SXS and $25.6~\mu$s for HXI and SGD). Therefore, the remaining 324 and 9~$\mu$s are distributed to other items in GPS-OFF and GPS-ON modes, respectively.
Such GPS modes only affect items~{\bf A} and {\bf B}; the error budget of item~{\bf A} is valid only in GPS-ON mode, and that of item~{\bf B} is defined by the GPS mode.
Item {\bf G} (the orbital determination accuracy) becomes worse in GPS-OFF mode than that in GPS-ON mode, but the budget include both modes.

The error budgets for items~{\bf A} and {\bf D} come from hardware design specification sheets of the GPSR and the SpaceWire user node (digital electronics; DE), respectively. 
The budget for item~{\bf B} in GPS-ON mode also comes from the hardware design, and that for GPS-OFF mode is defined as $270~\mu$s from the actual best-case performance of the Suzaku satellite~\cite{HXD}.
The budget for item~{\bf G} is the minimum requirement for the orbital determination group of the spacecraft operation. 
The remaining items~{\bf C} and {\bf E} are defined as 2.0 and 1.0~$\mu$s at maximum from a rough estimations from the system design and the software algorithm, respectively.

\section{Verification of Timing Performance}
\label{section:verification}
\subsection{Ground measurement of temperature dependence on quartz frequency}
\label{section:verification:suzaku}

As described in Section~\ref{section:design_software:ahmktim_ahtrendtemp}, temperature dependence of the quartz clock onboard SMU is described in the CALDB (quartz) and is used by {\it ahmktim} for time calculations in GPS-OFF mode.
The SMU has a function for self-measurement of this calibration item, but the operations are limited on orbit. 
Therefore, {\bf a)} detailed measurements of temperature dependence of the clock frequency and {\bf b)} functional tests of self-measurement in the flight configuration are performed on the ground before launch.

\begin{figure}[ht]
\centering
\includegraphics[width=0.75 \textwidth]{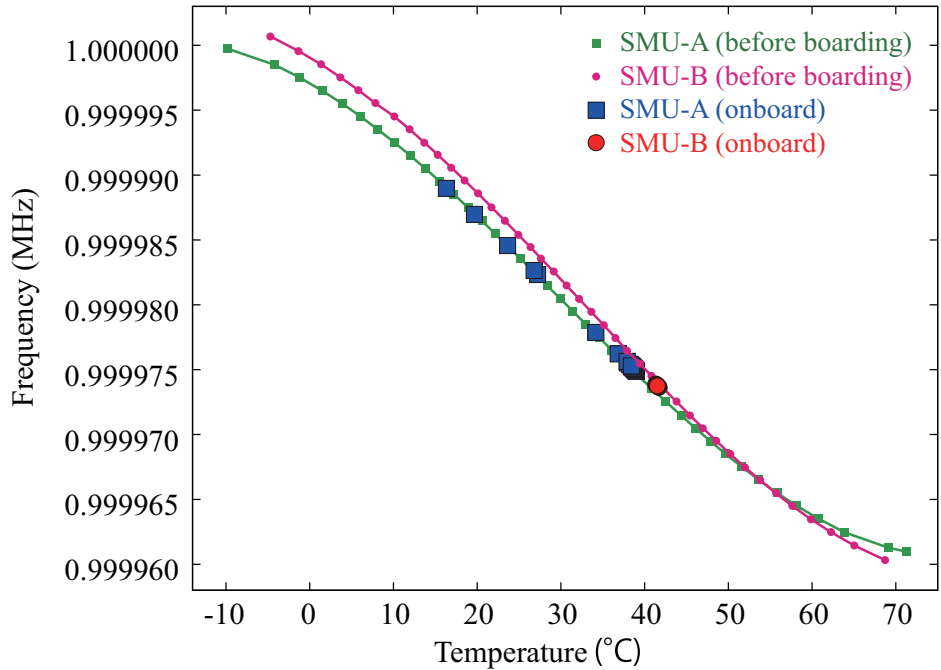}
\caption{Temperature dependence of the 1-MHz quartz-frequency in SMU-A and SMU-B
onboard the Hitomi satellite. Detailed measurements from before mounting on the spacecraft
are shown in green and magenta for SMU-A and SMU-B, respectively. 
Confirmation measurements during the thermal vacuum test on ground are shown in blue
and red for SMU-A and SMU-B, respectively.}
\label{fig:itemA_SMU_fvt_measurement}
\end{figure}

Detailed measurements of temperature dependence of the frequency of the quartz onboard the SMU (item~{\bf a}) were performed in October 2014, before boarding to the payload. 
Since the SMU has fully redundant components, two flight units (SMU-A and SMU-B) were used for the measurement.
The SMU was put in a vacuum heat bath and subjected to eight temperature cycles from $-10$ to +73 $^{\circ}$C at a rate of 1--2 $^{\circ}$C/min.
The quartz temperature was measured by a platinum temperature sensor at the electric board of the quartz.
The clock from the SMU quartz is available as a 1M~PPS signal from the SMU, and the frequency of this signal was measured by a frequency counter, which was well calibrated to 0.5~ppm accuracy. 
The frequency measurement was performed every 0.5~min. 
The relation between raw values for temperature and frequency show a small hysteresis between upward and downward progression of the temperature sweep at a 0.9\% level, so the measured points are averaged within a 1.0~Hz range when more than three points exist in that range.
Figure~\ref{fig:itemA_SMU_fvt_measurement} shows the results for SMU-A and SMU-B in green and magenta, respectively.
This tables are recorded in the CALDB (quartz) file for use by {\it ahmktim} (Fig.~\ref{fig:ahsoft}).

Functional test of self-measurement of the temperature dependence of the SMU quartz (item~{\bf b}) was performed in spacecraft thermal vacuum test from June to July 2015, when the spacecraft was in its flight configuration. 
During spacecraft thermal vacuum testing, hot and cold temperature environments were prepared for functional testing of onboard instruments. 
The self-measurement function of the temperature dependence of the SMU-A quartz was activated 30 times (for 16 s each) in GPS-ON mode during transition from hot to cold mode. 
The same function for SMU-B was also checked 9 times during cold mode.
Figure~\ref{fig:itemA_SMU_fvt_measurement} shows the results for SMU-A and SMU-B in blue and red, respectively. These results are consistent with the detailed measurement described above to within a 0.5~ppm level, which is sufficient for correction by {\it ahmktim}. 

In addition to items~{\bf a} and {\bf b} above, the temperature dependence can be roughly verified in the transition mode between the GPS-OFF to GPS-ON modes (see Section~\ref{section:design_software:ahmktim_ahtrendtemp}). This transition happened once on the ground during spacecraft thermal vacuum testing in July 2015 and once on orbit in February 2016.
On the ground, the GPSR mode changed from GPS-ON to GPS-OFF, then to GPS-ON again. The duration of GPS-OFF mode was 14,912~s and the temperature of the SMU-A was about 32.8~$^{\circ}$C, which corresponds to a 1PPS quartz frequency of about 0.9999797~Hz. 
Therefore, the free-run TI is expected to shift about 0.3034~s (19.4 ticks of L32TI) from TAI at the end of GPS-OFF mode, and 20.1 ticks of L32TI were actually observed, for consistency at a 4\% level. 
On orbit, the spacecraft operated in GPS-OFF mode from just after launch (17 Feb 2016) until when the GPS synchronization function was turned on (00:35 29 Feb 2016~UTC).
From 03:52:32 18~Feb 2016 UTC to 00:35:18 29~Feb 2016, L32TI shifted about 986.651 ticks, corresponding to 15.42~s.
During this period, which lasted about 938~ks, the SMU temperature was almost always about 26.3~$^{\circ}$C, at which the quartz frequency should be 0.9999839~Hz from Fig.~\ref{fig:itemA_SMU_fvt_measurement}, and thus about a 15.11~s shift is expected at this frequency over about 938~ks.
Therefore, the temperature dependence of the frequency of SMU-A quartz was also verified at about a 2\% level here. 

\subsection{Ground measurement of Item B: Delay from GPSR to SMU}
\label{section:verification:B_smu}

The measurement of item~{\bf B} in the error budget (Table~\ref{tbl:astroh_time_error_budget}) was performed during the first integration test in October 2013 when the GPSR, SMU, SpaceWire network, and SpaceWire user nodes were in their flight configuration. 
The goal was to measure the timing delay and jitter between the input and output of SMU, which are the GPSR 1PPS signal and the TIME\_CODE.
The former signal can be easily seen with an oscilloscope, but the latter is difficult to handle because the SpaceWire line is also used for communication of many commands and telemetries.
Therefore, a simple converter from the SpaceWire TIME\_CODE to a single digital pulse that is detectable by an oscilloscope was prepared for this measurement. Here, we call this converter the ``TIME\_CODE handler.'' 

Before main measurements in the flight configuration, the timing delay and jitter of the TIME\_CODE handler were measured using a commercial GPSR.
Measurements were performed over 25,169~trials under a SpaceWire link rate of 48~MHz at the same rate as the actual SMU SpaceWire port.
The results showed that the time delay becomes $387 \pm 6.4$~ns with 1-sigma error.
Note that, among this value, 219~ns were spent in recognizing the full signal length of TIME\_CODE by the TIME\_CODE handler, there was about a 2-ns delay due to SpaceWire cables, and that jitter by this GPSR is negligible (within 1~ns). 

\begin{figure}[ht]
\centering
\includegraphics[width=0.75 \textwidth]{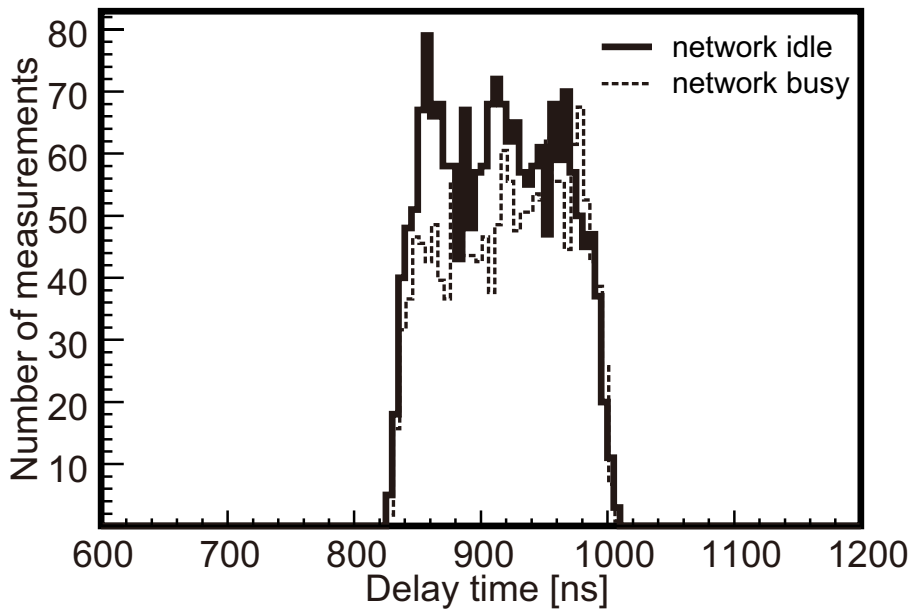}
\caption{Time delay within SMU from GPSR~1PPS signal to TIME\_CODE emitted from SMU. Intrinsic delay of the measurement system (387~ns) is included in the horizontal axis.}
\label{fig:itemB_SMU_delay_measurement}
\end{figure}

In main measurements of item~{\bf B}, the 1PPS signal from the GPSR was picked up from the communication line between SMU and GPSR, and the TIME\_CODE from the SMU was captured by the TIME\_CODE handler, pre-measured above, from the communication line between the SMU and the SpaceWire router in the network.
These two signals were detected by an oscilloscope, and the time delay and jitter between the two were measured.
The measurement is performed in two modes, when the SpaceWire network is unoccupied and when it is nearly fully occupied by communications between the components (hereinafter, these states are called ``idle'' and ``busy,'' respectively). 
In total, 1,910 and 1,619 measurements were performed in the idle and busy states, respectively.
Figure~\ref{fig:itemB_SMU_delay_measurement} shows the results.  In that figure, the horizontal axis includes the delay of the TIME\_CODE handler, 387~ns. 
Numerically, the results were $914 \pm 46$~ns and $920 \pm 47$~ns in the idle and busy states, respectively, with 1-sigma errors.
In summary, after subtracting the timing properties of the TIME\_CODE handler, the timing delay at SMU becomes $530 \pm 46$~ns.
The delay value of 530~ns is included in the CALDB (delay) file (see Fig.~\ref{fig:ahsoft}), which is used by the offline process with {\it ahtime}.
Since the jitter value for the error budget is defined as the 3-sigma, item~{\bf B} becomes about 137~ns, which satisfies the error budget (0.5~$\mu$s) in Table~\ref{tbl:astroh_time_error_budget}.

\subsection{Ground measurement of items C and D: Propagation time of TIME\_CODE}
\label{section:verification:CD_timecode}

\begin{table}[ht]
\renewcommand{\arraystretch}{1.3}
\caption{Delay and jitter of onboard components.}
\label{tbl:delay_components}
\centering
\begin{tabular}{lllll}
\hline 
ID & Component & Link rate & Delay & Jitter$^\dagger$\\
\hline 
a
& SMU
& 50 MHz
& 540 ns
& 143 ns\\
b 
& SpaceWire router
& 10 MHz
& 1814 ns
& 800 ns\\
c
& SpaceWire router
& 20 MHz
& 1114 ns
& 400 ns\\
d
& SpaceWire router
& 25 MHz
& 974 ns
& 320 ns\\
e
& SpaceWire router
& 50 MHz
& 694 ns
& 160 ns\\
f
& SpaceWire node FPGA
& 20 MHz
& 1600 ns
& 400 ns\\
g
& SpaceWire node CPU
& 20 MHz
& 1590 ns
& 390 ns\\
\hline 
\end{tabular}

$^\dagger$ Jitters are defined as the 3 times of standard deviations ($3 \sigma$). 
\end{table}

As described in Section~\ref{section:design_hardware:timing}, timing delays and jitter happen during propagation of the SpaceWire TIME\_CODE through the SpaceWire network tree from the SMU to user nodes. 
The jitter values correspond to item~{\bf C} in Table~\ref{tbl:astroh_time_error_budget} (see Section~\ref{section:time_uncertainty:items}), and all the delay values are listed in the CALDB (delay) file, which is used by the offline software {\it ahtime} (see Section~\ref{section:design_software:ahtime}).
Such delay and jitter values can be measured by probing many points in the SpaceWire network tree, but the number of points is limited in on-ground tests in the full satellite flight configuration due to the tight schedule before launch.
Therefore, delay and jitter values of each element in the SpaceWire network were first measured as summarized in Table~\ref{tbl:delay_components}, and the overall delay and jitter were estimated by a simple propagation simulator of TIME\_CODE under the SpaceWire network configuration (i.e., routing and the link rate) using pre-measured values of elements.
The overall delay time is just a linear sum of each delay (Table~\ref{tbl:delay_components}) in the routing path of the instruments listed in Table~\ref{tbl:delay_instruments}, but jitter is estimated by Monte Carlo simulation of the propagation of TIME\_CODE hops at each node.
Table~\ref{tbl:delay_instruments} shows the results of estimations of delay and jitter values.

\begin{table}[ht]
\renewcommand{\arraystretch}{1.3}
\caption{Delay and jitter estimated for instruments.}
\label{tbl:delay_instruments}
\centering
\begin{tabular}{llllcl}
\hline 
Instrument & Delay & Jitter & Source & Routing$^\dagger$ & Destination\\
\hline 
SXS
& 6092 ns
& 767 ns
  & GPSR & -a-e-e-d-f-g- & f\\
SXS-FW
& 3742 ns
& 270 ns
  & GPSR & -a-e-e-b- & SXS-FW\\
SXI
& 4502 ns
& 1041 ns
  & GPSR & -a-e-e-d-f- & f\\
HXI-1
& 3838 ns
& 755 ns
  & GPSR & -a-e-d-(cable)-f- & f\\
HXI-2
& 4505 ns
& 766 ns
  & GPSR & -a-e-e-d-(cable)-f- & f\\
CAMS
& 3048 ns
& 717 ns
& GPSR & -a-e-b- & f\\
SGD-1
& 3835 ns
& 476 ns
  & GPSR & -a-e-d-f- & f\\
SGD-2
& 4500 ns
& 493 ns
  & GPSR & -a-e-e-d-f- & f\\
\hline 
\end{tabular}

$^\dagger$ IDs (a,b,c,d,e,f,g) are defined in Table~\ref{tbl:delay_components}.
\end{table}

\begin{figure}[htb]
\centering
\includegraphics[width=0.55 \textwidth]{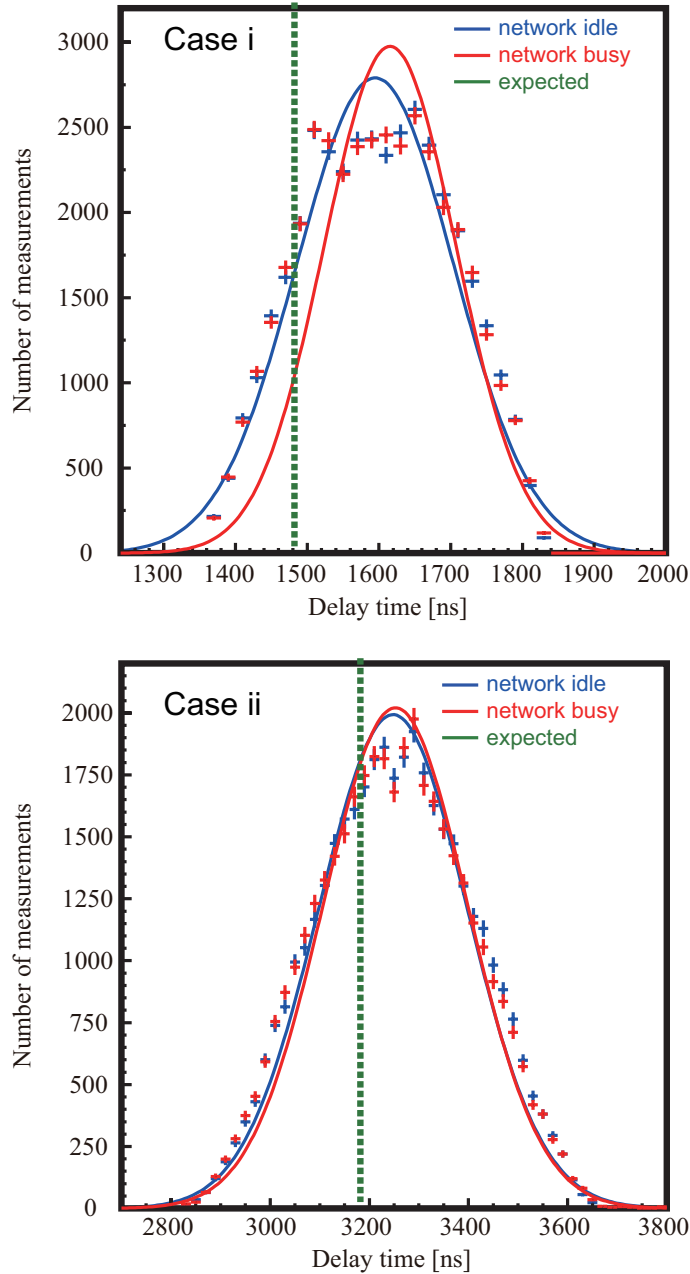}
\caption{Top and bottom panels represent propagation time distributions for cases {\bf i} and {\bf ii} in Table~\ref{tbl:delay_measurement}, respectively.
The blue and red crosses show the results for idle and busy networks, respectively.
The best-fit Gaussian functions are shown in the same color.
The green line shows the expected value by the propagation simulator (see the text).}
\label{fig:itemCD_SpWdelay_measurement}
\end{figure}

To verify the delay and jitter values estimated in Table~\ref{tbl:delay_instruments}, two paths from SMU to SGD-1 DE were selected and measured during the first integration test in October 2013.
Table~\ref{tbl:delay_measurement} lists the probe points, which are located at the point just after the SMU and before the CPU node or FPGA (field-programmable gate array) node of the SGD-1 instrument, which are assigned IDs of {\bf i} and {\bf ii}, respectively.
The propagation times for {\bf i} and {\bf ii} are estimated to be $1529 \pm 315$ and $3252 \pm 444$~ns, respectively, as listed in the table.
The SpaceWire MSR tester measured propagation times for TIME\_CODE in the idle and busy states 38,418 and 38,340 times, respectively.
Figure~\ref{fig:itemCD_SpWdelay_measurement} shows distributions of the propagation time in the two states.
The results show that network congestion does not affect propagation delay of the first-priority code TIME\_CODE at a 0.2\% level, and estimations of delay and jitter are consistent with actual measurements at less than a 10\% level.
The difference between estimated and measured values corresponds to item~{\bf E}, which turned out to be less than 10\% of the estimated delay time.
Numerically, it corresponds to 0.6~$\mu$s at maximum, which is within the error budget item~{\bf E}, 1~$\mu$s.

The two trials for delay and jitter values by propagation simulator were well verified, establishing the estimations in Table~\ref{tbl:delay_instruments}.
The delay values in that table were thus stored in the CALDB (delay) file and used by {\it ahtime}, and the jitter values in that table are all within the error budget item~{\bf C}, 2~$\mu$s.

\begin{table}[ht]
\renewcommand{\arraystretch}{1.3}
\caption{Measurement of SpaceWire delay times.}
\label{tbl:delay_measurement}
\centering
\begin{tabular}{lcccll}
\hline 
Id & Source&Routing$^\dagger$ & Destination     & Delay measured           & Delay expected \\
\hline 
{\bf i} & a & -x-e-d-x-g- & f(SGD-1) & $1595 \pm 330$ ns (idle) & $1529 \pm 315$ ns \\
   & &&                     & $1616 \pm 276$ ns (busy) &\\
{\bf ii} & a & -x-e-d-g-x- & f(SGD-1) & $3248 \pm 450$ ns (idle) & $3252 \pm 444$ ns \\
   &  &&                    & $3253 \pm 438$ ns (busy) &\\
\hline 
\end{tabular}

$\dagger$ a,b,c,d,e,f, and g are defined in Table~\ref{tbl:delay_components}. The 'x' indicates the measurement point.
\end{table}

\section{Summary of the Timing Performance}
\label{section:inorbit:summary}
\begin{table}[htb]
\renewcommand{\arraystretch}{1.3}
\caption{Summary of performance by the timing error budgets.}
\label{tbl:astroh_time_error_summary}
\centering
\begin{tabular}{llc}
\hline 
ID$^\dagger$ & Component & Performance\\
\hline 
A
& GPSR
& $0.01~\mu$s\\
B 
& SMU
& $0.14~\mu$s (GPS-ON)\\
C
& SpaceWire network
& 0.3 -- 1.0~$\mu$s (Table~\ref{tbl:delay_instruments})\\
D
& SpaceWire user node
& $<1.0~\mu$s\\
E
& Delay correction
& $<0.6~\mu$s\\
F
& Time resolution
& 5 or 25.6~$\mu$s (Table~\ref{tbl:summary_local_time})\\
G
& Orbit
& $0.3$ ns (GPS-ON), 0.5 $\mu$s (GPS-OFF)\\
\hline 
\end{tabular}

$^\dagger$ IDs (A, B, C, D, E, F, G) are defined in Table~\ref{tbl:astroh_time_error_budget}.
\end{table}

Table~\ref{tbl:astroh_time_error_summary} summarizes the performance of the Hitomi timing system, including the hardware and software designs for the seven items listed in Table~\ref{tbl:astroh_time_error_budget}.
Items~{\bf B}, {\bf C}, and {\bf D} were measured on the ground, as described in Sections~\ref{section:verification:B_smu} and \ref{section:verification:CD_timecode}, and item~{\bf F} is the time resolution of LOCAL\_TIME shown in Table\ref{tbl:summary_local_time}.
Items~{\bf A} and {\bf E} were designed following the specifications sheet.
Item~{\bf G} was estimated by the orbital determination group at JAXA after launch using the actual date on orbit. The position uncertainties determined by the GPSR are less than 1.7~m on orbit and a few centimeters by the offline determination process on the ground under the GPS-ON mode.
The accuracy becomes worse in the GPS-OFF mode when the orbital elements are determined by measurements of the position and velocity of the spacecraft from the ground station via ranging operations, but it is about 150~m.
Therefore, item~{\bf G} is negligible for the timing system both in GPS-ON and -OFF modes.
In summary, all seven items satisfy the error budget in Table~\ref{tbl:astroh_time_error_budget}.


\acknowledgments 
The authors would like to thank all the science and engineering members of the Hitomi team for their continuous contributions toward the development of instruments, software, and spacecraft operations.
This work was supported in part
by Grants-in-Aid for Scientific Research (B) from 
the Ministry of Education, Culture, Sports, Science and Technology (MEXT)
(No.~23340055 and No.~15H00773, Y.~T).

\newcommand{\procspie}{Proceedings of the SPIE}
\newcommand{\pasj}{Publications of the Astronomical Society of Japan}
\newcommand{\apj}{The Astrophysical Journal}
\newcommand{\aap}{Astronomy and Astrophysics}
\bibliographystyle{spiejour}   

\vspace{2ex}\noindent\textbf{ Yukikatsu Terada} is an associated professor at Saitama University. He received his BS and MS degrees in physics, and PhD degree in science from the University of Tokyo in 1997, 1999, and 2002, respectively.

\vspace{1ex}
\noindent Biographies and photographs of the other authors are not available.

\listoffigures
\listoftables

\end{spacing}
\end{document}